\documentclass[12pt]{article}
\usepackage{amssymb}
\usepackage{epsf}
\usepackage{latexsym}
\newcommand{\be}{\begin{equation}}
\newcommand{\ee}{\end{equation}}
\newcommand{\ba}{\begin{eqnarray}}
\newcommand{\ea}{\end{eqnarray}}

\begin{document}
\title{PP-waves from BPS Supergravity
Monopoles}
\author{D.H.~Correa$^a$\thanks{CONICET} \,,
E.F.~Moreno$^a$\thanks{Associated with CONICET} ,
S.~Reuillon$^{b,a}$
\\and \\
F.A.Schaposnik$^a$\thanks{Associated with CICPBA}\\
{\normalsize\it $^a$Departamento de F\'\i sica, Universidad Nacional
de La Plata}\\
{\normalsize\it C.C. 67, 1900 La Plata, Argentina}
\\
{\normalsize\it $^b$Laboratoire de Math\'ematiques et Physique Th\'eorique}\\
{\normalsize\it CNRS/UMR 6083, Universit\'e de Tours}\\
{\normalsize\it Parc de Grandmont, 37200 Tours, France}
}
\date{\hfill}
\maketitle

\begin{abstract}
We discuss the Penrose limit of the Chamseddine-Volkov BPS
selfgravitating monopole in four dimensional $N=4$ supergravity
theory with non-abelian gauge multiplets. We analyze the
properties of the resulting supersymmetric pp-wave solutions when
various Penrose limits are considered. Apart from the usual
rescaling of coordinates and fields we find that a rescaling of
the gauge coupling constant to zero is required, rendering the
theory abelian. We also study the Killing spinor equations showing
an enhancement of the supersymmetries preserved by the solutions
and discuss the embedding of the  pp-wave solution in $d=10$
dimensions.
\end{abstract}

Plane wave backgrounds arising in supergravity theories have
recently attracted much attention  in connection with the solution
of superstring theory and its relation with large N gauge theory
\cite{B1}-\cite{Cve2}. Such parallel plane (pp) wave solutions can
be obtained by taking the so-called Penrose limit \cite{Penrose}
adapted to the case of supergravity theories \cite{Gueven}. In the
present note we follow this approach to find pp-wave solutions in
$N=4$ gauged supergravity in 4 dimensions \cite{FS}, by starting
from the BPS selfgravitating monopole solution constructed by
Chamseddine and Volkov some time ago \cite{ChV1}-\cite{ChV2}. In
taking the Penrose limit, we had to adapt G\"uven treatment
\cite{Gueven}  to the case of gauged supergravity finding that not
only fields but also the gauge coupling constant should be scaled
in order to find consistent pp-wave solutions. In fact, the
coupling constant is rescaled to zero and the pp-wave solution
corresponds to an abelianized version of the theory. We consider
different limits and discuss their relation as a consequence of
the covariance property of the Penrose limit as originally
discussed in \cite{B3}. We also analize the uplifting of the
solution to 10-dimensional $N=1$ supergravity  and discuss the
issue of preserved supersymmetries.

We start by reviewing the $d=4$ gauged supergravity model and its
BPS monopole solution and then discuss the Penrose limit and the
resulting pp-wave configurations. Afterwards, we discuss how a
$d=10$ uplifted solution can be obtained from the $d=4$ theory by
an appropriate choice of the gauge group manifold.

~

The starting point is a four dimensional $ SU(2) \times SU(2)$
$N=4$ supergravity theory with non-abelian Yang-Mills multiplets
\cite{FS}. The field content of the theory includes, in the
bosonic sector, vierbein fields $e_\mu^a$, vector and
pseudo-vector non-abelian gauge fields $A_\mu^a$ and $B_\mu^a$
($a=1,2,3$), a dilaton $\phi$, and an axion {\bf a}. Concerning
the fermionic sector, there are 4 Majorana spin $3/2$ fields
$\psi_\mu^I$ and  4 Majorana spin $1/2$ fields $\chi^I$
($I=1,2,3,4$).

In the Chamseddine-Volkov (CV) monopole solution
\cite{ChV1}-\cite{ChV2} the pseudovector gauge field $B_\mu^a$ and
the axion {\bf a} are  put to zero. The bosonic part of the action
then reduces to
\be
S = \int  \sqrt{-g} \, d^4x \left( - \frac{1}{4} R +
\frac{1}{2} \partial_\mu
\phi \partial^\mu \phi - \frac{1}{4} \exp (2\phi) F_{\mu\nu}^a
F^{a \,\mu\nu}
+ \frac{e^2}{8} \exp(-2\phi)
\right)
\label{S1}
\ee
where
\be
F_{\mu\nu}^a = \partial_\mu A_\nu^a - \partial_\nu A_\mu^a +
e \varepsilon^{abc} A_\mu^b A_\nu^c
\ee

For purely bosonic configurations the supersymmetry transformation
laws reduce to
\ba
\delta \bar \chi  &=&  -\frac{i}{\sqrt 2} \bar \epsilon
\gamma^\mu \partial_\mu \phi - \frac{1}{2} \exp(\phi) \bar
\epsilon F_{\mu\nu}\sigma^{\mu\nu} + \frac{e}{4} \exp(-\phi)
\bar \epsilon \label{susy1}\\
\delta \bar \psi_\rho  &=&
\nabla_\rho \bar \epsilon  + \bar \epsilon \frac{e}{2} A_\rho  -
\frac{i}{2\sqrt 2}
\exp(\phi) \bar \epsilon F_{\mu \nu} \gamma_\rho \sigma^{\mu\nu}
+ \frac{i e}{4 \sqrt 2} \exp(-\phi)
\bar \epsilon \gamma_\rho
\label{susy2}
\ea
Here we have written $A_\mu = A_\mu ^a \alpha^a$, where
$\alpha^a_{IJ}$ are three of the six $4\times 4$ matrices which
generate the algebra of the original $SU(2)\times SU(2)$ symmetry
group (an explicit representation constructed in terms of the
Pauli matrices is given in \cite{FS}). We have also used
$\sigma^{\mu\nu}=\frac{1}{4}[\gamma^\mu,\gamma^\nu]$ and we have
written the 4 Majorana spinor parameters as
$\epsilon^I\equiv\epsilon$. $\nabla_\rho \bar \epsilon$ stands for
the spinorial covariant derivative
\be\label{cova}
\nabla_\rho \bar \epsilon = \partial_\rho \bar \epsilon
-\frac{1}{2} \bar \epsilon \omega_{\rho}^{\,\alpha \beta}
\sigma_{\alpha \beta}
\ee
where $\omega_{\rho}^{\, \alpha \beta}$
is the spin connection
(earlier Greek letters $\alpha, \beta, \ldots,$ correspond to the
locally flat system).

The spherically symmetric CV exact solution was found by
integrating the BPS equations that result from the vanishing of
the supersymmetry transformations (\ref{susy1})-(\ref{susy2}). The
explicit form of the metric and dilaton is given by
\ba
ds^2 &=& 2\exp(2 \phi)  \left(
dt^2 - \frac{1}{e^2}d\rho^2 - \frac{R^2(\rho)}{e^2} \left(d\theta^2
+\sin ^2\theta d\varphi^2 \right)\right) \label{met}\\
 \exp(2\phi) &=& a^2\frac{\sinh \rho}{2R(\rho)} \label{dilo} \\
R^2(\rho) &=& 2\rho \coth \rho - \frac{\rho^2}{\sinh^2\rho} - 1
\label{dil}
\ea
where $a$ is a free parameter reflecting scaling invariance
of the equations of motion.
Concerning the gauge field, the solution takes the form
\ba
\alpha^a A_\mu^a dx^\mu &=& \frac{w}{e} \left(-\alpha^2 d\theta +
\alpha^1 \sin \theta d\varphi\right)
+ \frac{1}{e}\alpha^3 \cos \theta d \varphi \nonumber\\
 w &=& \pm \frac{\rho}{\sinh \rho}
\label{AA}
\ea
Here $\rho$ is a radial-like  variable ($0 \leq \rho < \infty$)
implicitly defined  in terms of $w$ and $\phi$ while $a$ is a free
parameter which reflects the scale symmetry of Bogomol'nyi
equations. The geometry described by the metric (\ref{met}) is
everywhere regular and corresponds to a space whose topology is
$\mathbb{R}^4$. Concerning the gauge field solution, were $\rho$
the standard radial variable, it would exactly coincides with the
well-honored flat-space BPS gauge field solution
\cite{Bogo}-\cite{PS}, which corresponds to a charge $1$ magnetic
monopole. In the present case, being all gauge field massless, the
asymptotic behavior of $w$ in terms of the physical radial
variable $r$ is not exponentially decaying but
\be
w \sim \frac{\log r}{r^2} \;\;\; {\rm for} \;\;\;  r \to \infty
\label{ex}
\ee
As already noted in \cite{ChV1}-\cite{ChV2}, defining a magnetic
charge for this gauge field configuration is problematic since
there is no Higgs field breaking the symmetry and providing a
natural isospin direction to project the $SU(2)$ field strength on
the direction of the residual abelian symmetry (as one does for
the original flat space  't Hooft-Polyakov (tHP) monopole
configuration). Note however that ansatz (\ref{AA}) is nothing but
the gauge-transformed (with element $S$) of the original tHP
ansatz (we call $\bar A$ the gauge field ansatz in its original
tHP form) entangling space and isospace indices,
\ba
A &\to& \bar A = S^{-1}A S + \frac{i}{e} S^{-1} dS = \frac{i}{e} (1-w)
 [\Omega, d\Omega]\nonumber\\
\Omega &=& \frac{\alpha^a}{2} \frac{x^a}{r}
\label{nonumber}
\ea
Then, as for the pure Yang-Mills pioneering monopole ansatz of Wu
and Yang \cite{YW}, one can define a projected field strength in
the form
\be
\bar {\cal F}_{\mu\nu}  = {\rm Tr} \left(\bar F_{\mu\nu}  \Omega \right)
\ee
which, in the CV gauge becomes
\be {\cal F}_{\mu\nu}  = {\rm Tr} \left( F_{\mu\nu} S \Omega
S^{-1}\right) \label{cal} \ee
leading to a   magnetic field ${\cal B}_i = (1/2)
\varepsilon_{ijk}{\cal F}_{ij} $ which takes the form
\be
{\cal B}_r = (1 - w^2) \label{1-w}
\label{Br}
\ee
which indeed corresponds to a charge 1 magnetic monopole.

The Penrose limit procedure expands outwards the immediate
neighborhood of a null geodesic. Then, one has several
possibilities both in choosing the particular pair of variables
entering in the definition of the ``light-cone'' variables and how
one redefines and shifts the remaining variables. We shall explore
two in principle different Penrose limits of the solution
described above, and also discuss their relation.

The first Penrose limit is taken along a radial $\theta = \pi/2$
(and $\varphi = 0$) null geodesic. The appropriate change of
coordinates is in this case
\ba
t&=&\frac{1}{2}(u-\Omega^2v)\nonumber\\
\rho&=&\frac{e}{2}(u+\Omega^2v)\nonumber\\
\theta &=&\Omega y +\frac{\pi}{2}\nonumber\\
\varphi&=&\Omega x\
\label{name1}
\ea
where $\Omega$ is a positive real parameter.

The next stage corresponds to redefine the fields through an appropriate
scaling so that every term in the Lagrangian is scaled by the
same factor \cite{Gueven}. The fields redefinitions are the
following
\ba
\bar{g}_{\mu\nu}&=&\Omega^{-2}g_{\mu\nu} \nonumber\\
\bar{A}&=&\Omega^{-1}A.\
\label{noname1}
\ea
Moreover, to obtain a homogenous power of $\Omega$ in front of the
lagrangian, we should scale the gauge coupling constant
\ba
\bar{e}&=&\Omega e \, .\label{sce}
\ea
Let us perform this scaling on solution (\ref{met})-(\ref{AA}).
Now, the expressions of the new fields are
\ba
d\bar s^2 &=& -2\exp\left(2\phi\right)  \left(
dudv + \frac{R^2}{e^2} \left(dy^2
+\cos^2(\Omega y) dx^2\right)\right)
\nonumber\\
\exp(2\phi) &=& a^2
\frac{\sinh(\frac{e}{2}(u+\Omega^2v))}{2R(\frac{e}{2}(u+\Omega^2v))}
\nonumber\\
\alpha^a \bar A_\mu^a dx^\mu &=& \frac{w}{e} \left(-\alpha^2 dy
+ \alpha^1 \cos (\Omega y) dx\right)
-\frac{1}{e}\alpha^3 \sin(\Omega y) dx
\label{met2}
\ea

Then, the Penrose limit is accomplished by taking  $\Omega \to 0$.
It is worth noting that in this limit, the scaled coupling
constant (\ref{sce}) goes to zero and, as consequence, there is no
commutator in the field strengths. One can then consider the field
configuration obtained after the Penrose limit as a solution of an
abelianized version of the theory (\ref{S1}). In this sense the
Penrose limit relates solutions of a gauge supergravity theory to
solutions of a different gauge supergravity theory. The explicit
form of the solution is
\ba
d\bar s^2 &=& -2\exp\left(2\phi\right)  \left(
dudv + \frac{R^2}{e^2} \left(dy^2
+ dx^2\right)\right)
\label{met3a}\\
\exp(2\phi) &=& a^2\frac{\sinh (\frac{e}{2}u)}{2R(\frac{e}{2}u)}
\label{met3b} \\
\alpha^a \bar A_\mu^a dx^\mu &=& \frac{w}{e} \left( \alpha^1 dx
-\alpha^2 dy \right)
\label{met3}
\ea
It is straightforward to check that this pp-wave configuration,
 fulfills the equations of motion. Moreover,
as it is the case for the original CV solution,  it is everywhere
regular. Concerning  the pp-wave metric, it can be written in
Rosen form if one redefines the $u$ variable so that the dilaton
factor is eliminated from  $g_{uv}$. Concerning the gauge field
solution,  one can associate $A_\mu^1$ and $A_\mu^2$ with two
$U(1)$ gauge fields with electric and magnetic fields given by
\begin{eqnarray}
F_{tx}^1 &=& \frac{1}{e} \partial_t w\left(\frac{\rho +et}{2}\right)
\nonumber\\
F_{\rho x}^1 &=& \frac{1}{e} \partial_\rho w\left(\frac{\rho +
et}{2}\right) \label{campos1}
\end{eqnarray}
so that  $F_{tx}^1 =  e F_{\rho x}^1$, thus corresponding to a
plane wave travelling along the $\rho$ direction with orthogonal
electric and magnetic fields (For the other $U(1)$ gauge field one
gets an analogous result, with $y$  coordinate in place of $x$).
Notice that $F_{\mu\nu}^i F^{i\,\mu\nu} = 0$ and
$^*\!F_{\mu\nu}^iF^{i \,\mu\nu}= 0$ ($i=1,2$), that is, the
solution corresponds to a {\it null} field of the kind arising for
other plane-wave Einstein-Maxwell solutions. Let us comment that
if one uses the monopole gauge field configuration as in the
original 't Hooft-Polyakov ansatz instead of its gauge transformed
version (\ref{nonumber}), one arrives to the same pp-wave solution
(with the non-trivial gauge field components in a different
isospin direction).

As we stated above, we shall analyze a different Penrose limit
taken along  a radial $\theta = 0$  null geodesic. Of course this
geodesic belongs to the same orbit under the isometry group as
that corresponding to $\theta=\pi/2$. Then, as a consequence of
the covariance property of the Penrose limit introduced in
\cite{B3}, the solutions for the metrics should coincide. Now, in
the present case, it is interesting to analyze how the
corresponding gauge field solutions are connected. In fact, we
shall see that field configurations are related by a (singular)
gauge transformation.

The change of variables reads in  this case
\ba
t&=&\frac{1}{2}(u-\Omega^2v)\nonumber\\
\rho&=&\frac{e}{2}(u+\Omega^2v)\nonumber\\
\theta &=&\Omega r \nonumber\\
\varphi&=& \varphi
\label{name2}
\ea
The metric, gauge field and coupling constant scaling is the same
as in the previous case (eqs.(\ref{noname1})-(\ref{sce})). Proceeding
exactly as before we arrive, after taking the $\Omega \to 0$ limit,
to the following pp-wave solution,
\be d\bar s^2 = -2\exp\left(2\phi\right)  \left( du dv +
\frac{R^2}{e^2} \left(dr^2 + r^2d\varphi^2\right)\right)
\label{met3adife} \ee \be
 \exp(2\phi) = a^2\frac{\sinh (\frac{e}{2}u)}{2R(\frac{e}{2}u)}
\label{met3bdife}
\ee
\be
   \bar A_\varphi^1=  \frac{w}{e} r \;\;\;\;  \bar A_r^2 = -\frac{w}{e}
\label{met3dife}
\ee
Concerning the $\bar A_\varphi^3$ component, before taking the
$\Omega \to 0$ limit one has
\be
\bar A_\varphi^3 = \frac{1}{e\Omega} \cos \left(\Omega r\right)
\label{atres}
\ee
The corresponding nontrivial field strengths components are
\ba
\bar F_{t\varphi}^1 = \frac{1}{e} r \partial_t w \left (\frac{\rho
+ et}{2} \right) &\;\;\; & \bar F_{\rho\varphi}^1 = \frac{1}{e} r
\partial_\rho w \left (\frac{\rho + et}{2} \right)\nonumber\\
\bar F_{tr}^2 = -\frac{1}{e} \partial_t w \left (\frac{\rho +
et}{2} \right) &\;\;\; & \bar F_{\rho r}^2 =  -\frac{1}{e}
\partial_\rho w \left  (\frac{\rho + et}{2} \right)
\label{antesde}
\ea
Note that the $\Omega \to 0$ divergence in $\bar A_\varphi^3$ is
not harmful since this component has an associated field strength
$\bar F^3 = 0$ . Concerning its contribution to $\bar F^1$ and
$\bar F^2$,  the factor  $\bar e = \Omega e$ in front of
commutators  cancels out the singularity. Moreover, to order
$1/\Omega$  the $\bar A_\varphi^3$ component can be gauged out by
performing a (singular) gauge transformation
\be
g = \exp\left(\alpha_3 \frac{\varphi}{2} \right)
\ee
The gauge transformed field strength components read,
\ba
^g\bar F^1_{t\varphi} =\frac{1}{e} r \cos \varphi \partial_t w
&\;\;\; &
^g\bar F^1_{\rho\varphi} =\frac{1}{e} r \cos \varphi \partial_\rho w
\nonumber\\
^g\bar F^1_{tr} =-\frac{1}{e}  \sin \varphi \partial_t w
 &\;\;\; &
^g\bar F^1_{\rho r} =-\frac{1}{e}  \sin \varphi \partial_\rho w
\label{efes}
\ea
and analogous expressions for the $^g \bar F^2$ components.
As advanced these expressions for $ ^g\bar F^1$ (and the
corresponding one for $^g\bar F^2$), obtained by choosing the
Penrose limit according to eqs.(\ref{name2}) coincide with those
resulting from the alternative Penrose limit (\ref{name1}).
Indeed, equations (\ref{campos1}) and (\ref{efes}) coincide if one
writes in Cartesian coordinates the latter.

It is interesting to note that the field strength, written in the
form (\ref{antesde}), can be interpreted in terms of a
(gauge-dependent) magnetic field. Indeed, the $\bar F^1_{\rho
\varphi}$ component can be associated to a radial magnetic field
with flux
\be
\Phi = \frac{1}{2} \int \varepsilon_{ijk} \bar F^1_{jk}dS_i =
\frac{1}{e}\int_0^{2\pi} d\varphi
\int_{-et}^\infty d\rho \partial_\rho w\left(\frac{\rho + et}{2}\right)
 = \frac{4\pi}{e}
\ee
which corresponds to a unit   magnetic charge, as for   the
original CV monopole solution (see eq.(14)). Of course, as in this
last case, one is dealing with gauge dependent fluxes.

The analysis above on two isometric null geodesics leading to
 limiting solutions that should be identified, shows the relevance
 of classifying those Penrose limits that, in contrast, are inequivalent.
 This implies
 determining orbits of the isometry group of space time on
 the space of null geodesics. This has been done
 for spaces of the form $AdS \times S$ and supergravity brane solutions
 in \cite{B3}. For classifying the Penrose limit in the
   case of the CV monopole solution, one should start by observing that
 the resulting family of non-isometric Penrose
limits can be labelled by  the angular momentum  $l$ of the
corresponding null geodesic.  We have already presented above the
pp-waves configurations arising by taking the limit along radial
null geodesics ($l=0$). The classification should then be
completed by studying the Penrose limit along a generic null
geodesic.  The geodesic equations are in this case more involved
and deserve a careful analysis that we shall present elsewhere.

It is known that the Penrose limit can enhance the number of
preserved supersymmetries. We will see that this is the case for
the solution we are analyzing. Indeed, the original CV solution
preserves $1/4$ of the supersymmetries while, as we shall see, the
pp-wave solution eqs. (\ref{met3a})-(\ref{met3}) preserves 1/2. To
see this,  let us find the Killing spinors $\epsilon$ making
supersymmetry variations (\ref{susy1})-(\ref{susy2}) vanish. The
corresponding equations read
\ba
\delta\bar\chi\!\!&=& \!\!\frac{\exp(-\phi)}{\sqrt 2}\bar\epsilon
\gamma^{\underline u}
\left(-i\partial_u\phi
-
\left(\alpha^1\gamma^{\underline x}-\alpha^2\gamma^{\underline y}\right)
\frac{\partial_uw}{2R(u)}
\right)=0 \label{su1}
\\
\delta\bar\psi_\mu\!\!&=& \!\!
\nabla_\mu\bar \epsilon
-i \bar \epsilon\gamma_\mu \gamma^{\underline u}
\left(\alpha^1 \gamma^{\underline x}-\alpha^2\gamma^{\underline y}\right)
\exp(-\phi)\frac{\partial_uw}{4R(u)}=0 \label{su2}
\ea
where underlined indices in gamma matrices  refer to the tangent
space. Now, any spinor $\bar \chi$ can be written as
\be
\bar \chi =  \bar \chi^1 \gamma^{\underline v } +
\bar \chi^2 \gamma^{\underline u }
\ee
where
\be
\chi^1 =- \frac{1}{2} \gamma_{\underline u} \chi \; , \;\;\;
\chi^2 =- \frac{1}{2} \gamma_{\underline v} \chi
\ee
so that if we make the choice
\be
\bar \epsilon(u,v,x,y) =\bar \epsilon^1(u,v,x,y) \gamma^{\underline u}
\ee
eq.(\ref{su1}) is automatically satisfied. Concerning (\ref{su2}),
since $\omega_\mu^{\alpha\beta} \sigma_{\alpha\beta}\gamma^u =0$, for
$\mu=v,x,y$ it reduces
to
\be
\partial_\mu \bar \epsilon^1(u,v,x,y) \gamma^{\underline u}  = 0
\ee
so that $\epsilon^1=\epsilon^1(u)$, whereas the remaining equation reads
\be
\left(\partial_u \bar \epsilon^1(u)
+i\bar \epsilon^1(u) \left(\alpha^1\gamma^{\underline x}-
\alpha^2\gamma^{\underline y}\right)\frac{\partial_uw}{2R(u)}
\right)\gamma^{\underline u}= 0
\ee
which can be easily integrated. We then conclude that at least
the pp-wave solution preserves $1/2$ of the 16 supersymmetries of
the theory. To determine whether there are additional Killing
spinors, one should look for spinors of the form $\bar
\epsilon^2(u,v,x,y) \gamma^{\underline v}$ with vanishing
supersymmetry variation. But then, from eq.(\ref{su1}) one
concludes that the following equation should hold,
\be
\bar \epsilon^2\gamma^{\underline v}\gamma^{\underline u}
\left(-i\partial_u\phi
-
\left(\alpha^1\gamma^{\underline x}-\alpha^2\gamma^{\underline y}\right)
\frac{\partial_uw}{2R(u)}
\right)=0
\ee
Now, in view of the explicit $u$-dependence of $\phi$, $R$  and
$w$ given in (\ref{dilo})-(\ref{AA}), the only possibility is that
$\epsilon^2 = 0$. In conclusion, one only finds the standard
Killing spinors and hence the supersymmetries preserved by the
pp-wave are  1/2, that is, twice those preserved by CV solution.

Four dimensional  solutions like the original monopole solution
(\ref{met})-(\ref{AA}) or the pp-wave solution
(\ref{met3a})-(\ref{met3}) can be uplifted to $d=10$ dimensions as
solutions of $N=1$ supergravity. Concerning the former, this was
done in \cite{ChV2} and the same procedure can be used to uplift
the latter. We shall here briefly describe the last uplifting.

Following \cite{ChV2}, we shall use a hat   to distinguish fields
in ten dimensions and Latin capital letters refer to the
coordinates. The sets $(M,N,P, \cdots)$ and $(A,B,C, \cdots)$
stand for curved space and tangent space indices respectively.
These sets can be decomposed into space-time and internal indices
from the 4-dimensional viewpoint. That is, $\{M\}=\{\mu,m\}$ and
$\{A\}=\{\alpha,{\rm a}\}$ ($\mu, \alpha = 0,1,2,3$ and $m, {\rm
a} = 4,5,\ldots,9$).

The action for the $N=1$ supergravity in $d=10$ is
\be
\hat S = \int  \sqrt{-\hat g} \, d^{10}x
\left( - \frac{1}{4} \hat R + \frac{1}{2} \partial_\mu
\hat \phi \partial^\mu \hat\phi + \frac{1}{12} \exp (-2\hat\phi)
\hat H_{MNP}\hat H^{MNP}
\right)
\label{S10}
\ee
where the antisymmetric field strength tensor is given in terms of
a two-form gauge field
\be
\hat H_{MNP}=\partial_M\hat B_{NP}+\partial_N\hat B_{PM}
+\partial_P\hat B_{MN}
\ee

The reduction from $d=10$ to $d=4$ can be done in many different
ways. Depending on the six-dimensional manifold ${\cal M}_6$  used
to compactify the original $d=10$   space to $d=4$ space one
obtains different $d=4$ gauged supergravity. For instance,
compactifying with $S^3\times S^3$ or $S^3\times T^3$ the
resulting supergravities in $d=4$ are gauged by $SU(2)\times
SU(2)$ or $SU(2)\times [U(1)]^3$ respectively.

If the manifold spanned by $\{x^m \}$ forms a compact group space,
there will be functions satisfying
\be
(U^{-1})_{\rm b}^{\ m}(U^{-1})_{\rm c}^{\ n}(\partial_m U^{\rm a}_{\ n}-\partial_n U^{\rm a}_{\ m})=
\frac{f_{\rm abc}}{\sqrt 2}
\ee

Let us present the ansatz for expressing the 10-dimensional fields
in terms of their 4-dimensional counterparts and the functions
$U^{\rm a}_{\ m}$. The 10-dimensional metric tensor is
\ba\label{upg}
\hat g_{\mu\nu}&=&\exp(-3\phi/2)g_{\mu\nu} -2\exp(\phi/2)
A_\mu^{\rm a} A_\nu^{\rm a}\nonumber\\
\hat g_{\mu m}&=&\sqrt 2\exp(\phi/2)A_\mu^{\rm a}
U^{\rm a}_{\ m}\nonumber\\
\hat g_{mn}&=&-\exp(\phi/2) U^{\rm a}_{\ m}U^{\rm a}_{\ n}
\ea
The ansatz for the dilaton is simply
\be\label{upfi}
\hat\phi=-\frac{\phi}{2}
\ee
Finally, the 10-dimensional field strength tensor takes the form
\ba
\hat H_{\alpha\beta\gamma} &=& \exp(-7\phi/4)\
\varepsilon_{\delta\alpha\beta\gamma}\partial^\delta{\mathbf{a}}
\label{upH}\\
\hat H_{\alpha\beta{\rm a}} &=& -\frac{1}{\sqrt 2}\exp(5\phi/4)
F^{\rm a}_{\alpha\beta}\label{upH2}\\
\hat H_{\alpha{\rm ab}} &=& 0 \label{upH3}\\
\hat H_{{\rm abc}} &=& \frac{1}{2\sqrt 2}\exp(-3\phi/4)f_{\rm abc}
\label{upH4}
\ea

Some comments are in order. The field {\bf a}, appearing in
eq.(\ref{upH}) is the axion, which is set to zero in the solutions
that we will uplift. The field strength in (\ref{upH2}) takes
values in the Lie algebra of the six dimensional manifold

\be
F^{\rm a}_{\alpha\beta}= \partial_\alpha A_\beta^{\rm a} -
\partial_\beta A_\alpha^{\rm a} +
f_{\rm abc} A_\alpha^{\rm b} A_\beta^{\rm c}
\ee
%

We shall now proceed to construct a pp-wave solution to the
equations of motion of the $d=10$ supergravity model. As stated
above, one can directly uplift to this end, the  $d=4$ pp-wave
solution (\ref{met3a})-(\ref{met3}).

The pp-wave solution (\ref{met3a})-(\ref{met3}) involves just two
gauge field components $A_\mu^1$ and $A_\mu^2$ which, being the
gauge coupling constant zero in the Penrose limit, lead to two
abelian field strengths. It is then natural to choose the six
dimensional group manifold so as to arrive to a model with just
abelian gauge fields, namely ${\cal M}_6 = T^6$. The appropriate
choice of functions $U^{\rm a}_{\, m}$  is
\be
U^{\rm a}_{\, m} = \delta^{\rm a }_{\, m}
\label{udelta}
\ee
This choice, together with the explicit form for $A_\mu^1$ and
$A_\mu^2$ given in (\ref{met3a})-(\ref{met3}), leads to the
following ten-dimensional pp-wave metric,
\begin{eqnarray}
\hat g_{\mu\nu} &=& \exp(-3\phi/2) g_{\mu\nu} -2 \exp(\phi/2)
\frac{w^2}{e^2}
\left( \delta_\mu^{x} \delta_\nu^{x} +
\delta_\mu^{y} \delta_\nu^{y}
\right) \nonumber\\
\hat g_{x 4} & = & \sqrt 2 \exp(\phi/2) \frac{w}{e} \nonumber\\
\hat g_{y 5} & = & -\sqrt 2 \exp(\phi/2) \frac{w}{e} \nonumber\\
\hat g_{mn} &=& - \exp(\phi/2) \delta_{mn}
\label{met10}
\end{eqnarray}
Concerning the field strength components, the only non-vanishing
ones take the form
\begin{eqnarray}
\hat H_{u x 4} &=& -\frac{1}{\sqrt 2e} \, \frac{dw}{du}\nonumber\\
\hat H_{u  y  5} &=&  \frac{1}{\sqrt 2e} \, \frac{dw}{du}
\label{Hdiez}
\end{eqnarray}
Finally the solution for the dilaton is given by
\be
\exp (4\hat \phi )= \exp (2 \phi) =
a^2\frac{\sinh (\frac{e}{2}u)}{2R(\frac{e}{2}u)}
\label{dilaton}
\ee
We have checked, after a straightforward but tedious computation,
that the pp-wave configuration (\ref{met10})-(\ref{dilaton}) is a
solution of the $d=10$ $N=1$ supergravity equations of motion.

In summary, we have studied the Penrose limit of the
Chamseddine-Volkov BPS monopole solution to the equations of
motion of $N=4$ supergravity theory with non-abelian gauge
multiplets in $d=4$ dimensions. We found that in order to define a
consistent Penrose limit, the gauge coupling constant should be
scaled to zero, making the limit theory abelian. Then, contrary to
what happens in ungauged supergravity theories, the Penrose limit
relates two different gauge supergravity theories.

We have analyzed the properties of the resulting pp-wave solution
showing that an enhancement of supersymmetry takes place as it
usually occurs when a supergravity solution is transformed to a
plane wave geometry. We have also discussed the uplifting of the
pp-wave configuration  so that it becomes a solution of $N=1$
supergravity in $d=10$ dimensions. In this respect, let us recall
that, apart from its intrinsic interest, the CV solution proved to
be very useful in finding a smooth solution of the seven
dimensional supergravity model that was analyzed  by Maldacena and
N\'u\~nez in their study of the large $N$ limit of pure ${\cal
N}=1$ Super Yang-Mills theory \cite{MN}. The corresponding Penrose
limit was studied in \cite{GO} where it was shown that also a
supersymmetry enhancement takes place as we have shown it happens
for the original CV solution. It would be interesting to explore
this enhancement  in the uplifted solution as well as its
consequences for the related gauge theories. We hope to analyze
this issues in a forthcoming work.

\vspace{1 cm}

\noindent\underline{Acknowledgements}: We thank Carlos N\'u\~nez,
Peter Forgacs, Mart\'\i n Schvellinger and Mikhail Volkov   for
very useful comments and insights. S.R. acknowledges the
Universidad de La Plata for hospitality. This work  was partially
supported by UNLP, CICBA, CONICET, ANPCYT (PICT grant 03-05179)
Argentina and ECOS-Sud Argentina-France collaboration (grant
A01E02).  E.F.M. is partially supported by Fundaci\'on Antorchas,
Argentina.

%

\end{document}